\documentstyle[preprint,tighten,eqsecnum,aps]{revtex}
\begin{document}
\preprint{TRI-PP-94-90}
\draft
\title{$\Delta$--Excitation and Exchange Corrections for NN--Bremsstrahlung }
\author{M.\ Jetter and H.\ W.\ Fearing}
\address{TRIUMF, 4004 Wesbrook Mall, Vancouver, B.\ C., \\
 Canada  V6T 2A3}
\date{October 19, 1994}
\maketitle
\begin{abstract}
The role of the relativistic amplitudes for a number of ${\cal O}(k)$
processes usually neglected in potential model calculations of
NN--bremsstrah\-lung is investigated. In particular, we consider the
$\Delta$--excitation pole contributions related to the one--pion and
one--rho exchange and in addition include the exchange contributions
induced by the radiative $\omega,\,\rho \to \pi \gamma$
decays. The contributions are calculated from relativistic Born
amplitudes fitted to $\Delta$--production and absorption data
in the energy range up to 1 GeV and then
used to supplement potential model and soft photon calculations
for nucleon--nucleon bremsstrahlung. The effects on
$NN\gamma$--observables, although moderate in general, are found to be
important in some kinematic domains.
\end{abstract}
\narrowtext
\section{Introduction}

Nucleon--nucleon bremsstrahlung has been extensively investigated both
experimentally and theoretically during the past 30 years. For $pp \to
pp\gamma$
the experimental data available covers an energy domain between 42 MeV
\cite{Jov80} and 730 MeV \cite{Nef77}, the most recent results being
obtained at incident proton energies around the pion production
threshold \cite{Mic90,Kit86,Prz92}. Up to the pion production
threshold, the $NN\gamma$ potential model
using realistic NN--potentials for the
description of the nuclear force plus first order approximation for the
electromagnetic interaction gives the most
successful description.

A comparison of different NN--interactions yields
widely equivalent $NN\gamma$--results in the whole kinematic range
\cite{Jet94}, so that bremsstrahlung provides a sensitive test for the
dynamical model used to describe the photon emission.
In the framework of the potential model, however, there is
little controversy about the basic features and the
correction terms to be used. In particular, relativistic
spin corrections as derived in  \cite{Wor86,Lio72}
and rescattering contributions  \cite{Nak89}
are used in the most recent analyses \cite{Jet94,Her92},
see also \cite{Bro91,Ari93}. Moreover, the role of Coulomb
corrections in $pp\gamma$ \cite{Ari94} and exchange currents in the
$np\gamma$ cross section have been discussed \cite{Nak89,Bro73}.

The main theoretical problems left are presumably related to
shortcomings in principle of the potential model description.
For example, Lorentz invariance can only be accounted for
approximately by a covariant treatment of the kinematic transformations,
inclusion of higher order terms in $(p/m)$ in the electromagnetic
operator (relativistic spin corrections) and an appropriate description
of the NN--interaction. Further common features of existing potential
model calculations are the absence of dynamical baryon resonances and
the neglect of two--body currents beyond the ${\cal O}(k^0)$ terms
given by the soft photon approximation (SPA).

The role of the $\Delta$--excitation was studied in the 1970's in
two different approaches. The authors of \cite{Boh77} used a dispersion
analysis in order to correct the one--pion exchange neutron--proton
electromagnetic current. The effect of the $\Delta$--resonance
on the $np\gamma$ coss section at 200 MeV turned out to be small.
In \cite{Szy77} and \cite{Tia78}, the $\Delta$--excitation part of
the $pp\gamma$ amplitude was derived from phenomenological
Lagrangians and combined with soft photon or one--boson exchange
Born calculations for the radiative background.

A more sophisticated potential approach
including effects of the $\Delta$ based on a coupled channel
calculation of the half off--shell NN-- and N$\Delta$ T--matrices
together with a phenomenological $N \Delta \gamma$--vertex has been published
recently \cite{Jon93,Ede94}. In contrast to the $np\gamma$--results of
\cite{Boh77}, the authors find an appreciable $\Delta$ contribution
to the 280 MeV $pp\gamma$ observables.

For higher energies where the potential model is inappropriate,
the closely related problem of dilepton production has recently been
studied in the framework of an effective one--boson exchange plus
$\Delta$--excitation Born approximation \cite{Sch94}.

As mentioned above, a rigorous derivation of induced two--body currents
is not possible to all orders. The reason is that, unlike e.\ g.\
for a one-pion exchange potential \cite{Ris89}, the gauge invariant
replacement $V_N(\vec p) \to V_N\left( \vec p - e \vec A \right)$
in the argument of a general NN--potential $V_N$  \cite{Nak89,Bro73}
leads to an unique expression for the induced current only in SPA.
To this order, the induced current approximates contributions due to
the exchange of charged mesons and vanishes for $pp\gamma$. The radiative
decay processes we are considering here are thus not accounted for in
the conventional potential model. So far, only the radiative
$\omega \to \pi^0 \gamma$ decay for $pp\gamma$ has been considered
as an example of such an ${\cal O}(k)$ internal radiation process
\cite{Kam77}.

The purpose of the present work is to provide an estimate of the role of
baryon resonances and internal radiation processes reliable up to
photon energies of about 300 MeV corresponding to the highest photon
energy possible in the 730 MeV $pp\gamma$--experiment of \cite{Nef77}.
{}From a comparison with recent analyses of pion--photoproduction \cite{Gar93},
we expect the $\Delta(1232)$--resonance and the radiative decay of the
$\omega$ and $\rho$ to be the leading corrections.

A coupled channel calculation for the $NN - N\Delta$ system omitting
the contribution of the $\Delta  \Delta$--states \cite{Ter87} is typically
represented by a set of Lippmann--Schwinger equations
\begin{eqnarray}
 T_{NN} & = & V_{NN} + V_{NN} G_N T_{NN} + V_{N\Delta} G_{\Delta}
T_{\Delta N}, \label{lip1} \\
T_{\Delta N} & = & V_{\Delta N} + V_{\Delta N} G_N T_{NN} +
 V_{\Delta \Delta} G_{\Delta} T_{\Delta N}. \label{lip2}
\end{eqnarray}
Inclusion of the $\Delta$ thus modifies the NN--amplitude $T_{NN}$
and yields an additional amplitude $T_{\Delta N}$. The pure
NN--interaction below the pion production threshold is known
to be well described by phenomenological and/or meson theoretic potentials
so that we feel safe identifying $T_{NN}$ of equation (\ref{lip1})
with a pure NN-T-matrix calculated from a realistic nuclear potential
in effect absorbing the $V_{N \Delta} G_{\Delta} T_{\Delta N}$ piece.
By then adding the $\Delta$--excitation Born terms, we basically
neglect the modifications of the amplitude $T_{\Delta N}$ (equation
(\ref{lip2})) induced by the iterative terms.
Empirically, i.\ e.\ from a comparison of the Born amplitude $V_{\Delta N}$
with coupled channel $\Delta$--absorption predictions
for lab energies of $50 \, - 300$ MeV \cite{Ter87}, and experimental
$\Delta$--production data at
800 MeV \cite{Hud78} and 970 MeV \cite{Dmi86}, it turns out that the
effect of the iteration can be simulated by a simple energy--dependent
rescaling of the Born amplitude in good approximation. We are then left
basically with possible errors originating from the superposition of a
(necessarily real and therefore not unitary) Born amplitude with
the iterated potential model and SPA amplitudes.
Our approach for the $\Delta$ is thus complementary to a full
coupled channel calculation \cite{Jon93} which systematically avoids
the shortcomings mentioned above but cannot be extended easily to
processes such as the internal radiative decays and -- at least in
the $NN\gamma$--case -- cannot be applied at higher energies.

In the following two sections, a description of the model together
with a discussion of the parameters will be given.
Results for $pp\gamma$ and $np\gamma$ are presented in section IV.
We have taken care to test thoroughly the influence of the
experimental and theoretical uncertainties for the various
ingredients of the amplitude that can not be fixed to accurate data.
The results might be useful to estimate the reliability not only of
this but of any comparable calculation.

\section{Inclusion of the $\Delta$}

For the derivation of the $\Delta$--excitation terms depicted in Fig. 1
we start from the interaction Lagrangians for the isovector mesons
$\pi$ and $\rho$ (see \cite{Mac89}) in the notation of \cite{Bjo64}:
\begin{eqnarray}
{\cal L}_{\pi \scriptscriptstyle NN}
 &=& -ig_{\pi \scriptscriptstyle NN} \bar \psi  \gamma_5 \vec \tau
\psi \vec \pi \label{lag1} \\
{\cal L}_{\rho \scriptscriptstyle  NN} &=& -g_{\rho
 \scriptscriptstyle NN}\left(  \bar \psi  \gamma^{\mu}
 \vec \tau \psi \vec \rho_{\mu} + {\kappa_{\rho} \over 4
 m_{\scriptscriptstyle N }}
\bar \psi \sigma^{\mu \nu} \vec \tau \psi (\partial_{\mu} \vec \rho_{\nu}
- \partial_{\nu} \vec \rho_{\mu}) \right) \label{lag2} \\
{\cal L}_{\pi \scriptscriptstyle N \Delta} &=&
 -{g_{\pi \scriptscriptstyle N\Delta} \over m_{\pi}
} \bar \psi \vec T \psi_{\Delta}^{\mu} \partial_{\mu} \vec \pi + H.c.
\label{lag3}  \\
{\cal L}_{\rho \scriptscriptstyle N \Delta} &=&
 i {g_{\rho \scriptscriptstyle N\Delta} \over m_{\rho}}
 \bar \psi \gamma^5 \gamma^{\mu}
\vec T \psi_{\Delta}^{\nu} (\partial_{\mu} \vec \rho_{\nu}
-\partial_{\nu} \vec \rho_{\mu})  + H.c.\, , \label{lag4} \end{eqnarray}
where $\psi_{\Delta}$ is the spin--3/2 Rarita--Schwinger field and $\vec T$
stands for the isospin operator  for the transition of an isospin--1/2
and isospin--1 to an isospin--3/2 particle \cite{Mac89}. We have used the
pseudscalar $\pi NN$--coupling for convenience. As the $\pi NN$--vertices
in our calculations involve only on--shell nucleons, pseudovector coupling
for the pion instead of equation (\ref{lag1}) would not change any of
our results.
{}From eq. (\ref{lag1}) -- (\ref{lag4}), the vertex functions read
\begin{eqnarray}  \Lambda_{\pi \scriptscriptstyle NN} &=&
g_{\pi \scriptscriptstyle NN} \gamma_5, \qquad
\Lambda_{\rho \scriptscriptstyle NN}^{\mu}(q) =
- i g_{\rho \scriptscriptstyle NN} \left( \gamma^{\mu} - {\kappa_{\rho}
\over 2 m_{\scriptscriptstyle N}} i \sigma^{\mu \nu}
q_{\nu} \right), \nonumber \\
\Lambda_{\pi \scriptscriptstyle N\Delta}^{\mu}(q)& =&
 {g_{\pi \scriptscriptstyle N\Delta} \over m_{\pi}} q^{\mu},
\qquad \Lambda_{\rho \scriptscriptstyle N\Delta}^{\mu \nu}(q) =
 i {g_{\rho \scriptscriptstyle N\Delta} \over m_{\rho}
} \left( \not q g^{\mu \nu} - \gamma^{\mu} q^{\nu} \right) \gamma_5
\label{vertices} \end{eqnarray}
and one derives the transition amplitudes
for diagram (a) in Fig. 1 \cite{Tia78}:
\begin{eqnarray} T^{(1)}(p_1,p_2;p_3,p_4)
&=&  \bar u (p_3) \Lambda^{\nu}_{\pi \scriptscriptstyle N \Delta}
(q) P^{\Delta}_{\nu \sigma} (p_1-k)
\epsilon_{\mu} \Gamma^{\mu \sigma}_{\gamma \scriptscriptstyle N \Delta}
(k,p_1-k) u(p_1)
\nonumber \\ \mbox{} && \quad
 P^{\pi}(q) \bar u(p_4) \Lambda_{\pi \scriptscriptstyle NN} u(p_2)  \nonumber
\\  \mbox{} && + \bar u (p_3)
 \Lambda^{\lambda \nu}_{\rho \scriptscriptstyle N \Delta }
(q) P^{\Delta}_{\nu \sigma} (p_1-k)
\epsilon_{\mu}  \Gamma^{\mu \sigma}_{\gamma \scriptscriptstyle N \Delta}
(k,p_1-k) u(p_1)
\nonumber \\ \mbox{} && \quad
 P^{\rho}_{\lambda \tau}(q) \bar u(p_4)
 \Lambda^{\tau}_{\rho \scriptscriptstyle NN} u(p_2)
 ,\label{diaga}  \end{eqnarray}
and equivalently for diagram (b)
\begin{eqnarray} T^{(2)}(p_1,p_2;p_3,p_4)
&=&  \bar u (p_3)
\epsilon_{\mu} \tilde \Gamma^{\mu \sigma}_{\gamma
\scriptscriptstyle N \Delta} (k,p_3+k)
 P^{\Delta}_{\sigma \nu} (p_3+k)
\tilde \Lambda^{\nu}_{\pi \scriptscriptstyle N \Delta} (q)
u(p_1)
\nonumber \\ \mbox{} && \quad
 P^{\pi}(q) \bar u(p_4) \Lambda_{\pi \scriptscriptstyle NN} u(p_2)  \nonumber
\\  \mbox{} && + \bar u (p_3)
\epsilon_{\mu} \tilde \Gamma^{\mu \sigma}_{\gamma
\scriptscriptstyle N \Delta} (k,p_3+k)
 P^{\Delta}_{\sigma \nu} (p_3+k)
\tilde \Lambda^{\lambda \nu}_{\rho \scriptscriptstyle N \Delta} (q)
 u(p_1)
\nonumber \\ \mbox{} && \quad
 P^{\rho}_{\lambda \tau}(q) \bar u(p_4)
 \Lambda^{\tau}_{\rho \scriptscriptstyle NN} u(p_2) .
 \label{diagb}  \end{eqnarray}
Here, the energies and momenta are constrained according to
$p_1+p_2=p_3+p_4+k$, $q=p_4-p_2$ is the four momentum
transferred by the meson,
$\epsilon_{\mu} $ the photon unit polarization vector, $u, \,
\bar u$ denote free nucleon Dirac spinors and $P$ the appropriate propagators
for the mesons and the $\Delta$. With the tilde
in equation (\ref{diagb}) we indicate that the vertex function
has to be taken from the Hermitian conjugate of the corresponding
Lagrangian.

Associated with eqs. (\ref{diaga}) and (\ref{diagb}) are isospin factors,
e.\ g.\ in case of proton--proton bremsstrahlung:
$$\langle \chi_p \mid T_3 \mid \chi_{\scriptscriptstyle \Delta^+} \rangle
\langle \chi_{\scriptscriptstyle \Delta^+} \mid T_3 \mid \chi_p \rangle
\langle \chi_p \mid \tau_3 \mid \chi_p \rangle = {2 \over 3} \nonumber
$$
 The full $\Delta$ excitation amplitude including the isospin factors
and a relative minus for graphs involving interchange of the final state
particles is then
\begin{eqnarray} T_{\Delta}^{pp\gamma} &=& {2 \over 3} \sum_{i=1}^2
\left\{ T^{(i)}(p_1,p_2;p_3,p_4) + T^{(i)}(p_2,p_1;p_4,p_3)
\right. \nonumber \\  && \mbox{} - \left.
 ( T^{(i)}(p_1,p_2;p_4,p_3) + T^{(i)}(p_2,p_1;p_3,p_4)  ) \right\}
 \nonumber \\
T_{\Delta}^{np\gamma} &=&  {2 \over 3}  \sum_{i=1}^2
\left\{
( - T^{(i)}(p_1,p_2;p_3,p_4) + T^{(i)}(p_2,p_1;p_4,p_3) )
\right. \nonumber \\ && \mbox{}  + (-1)^{(i)} \left.
( - T^{(i)}(p_1,p_2;p_4,p_3) +  T^{(i)}(p_2,p_1;p_3,p_4))
\right\}. \label{deltamp} \end{eqnarray}

For the electromagnetic $N\Delta$ current of diagram (a)
we use the parametrization \cite{Jon73}
\begin{equation} \Gamma_{\gamma \scriptscriptstyle N\Delta }^{\mu \nu}
 (k,p)= -ie
\left\{ {G_1 \over m_{\scriptscriptstyle N}} (\gamma{^\mu} k^{\nu}
- \not k g^{\mu \nu}) +
{G_2 \over m_{\scriptscriptstyle N}^2}
 (p{^\mu} k^{\nu} - k \cdot p \, g^{\mu \nu} ) \right\}
\gamma_5\, . \label{gamdel} \end{equation}
$\tilde \Gamma_{\gamma \scriptscriptstyle N \Delta}(k,p)$
is equal to $\Gamma_{\gamma \scriptscriptstyle N \Delta}(k,p)$
 up to a relative minus sign in the
term proportional to $G_1$. The couplings $G_1$ and $G_2$ are
experimentally determined by a fit to the $M1^+/E1^+$ multipole
pion--photoproduction cross section. Values for $G_1$ and $G_2$
range from $G_1=2.0$, $G_2=0$, the prediction of the vector dominance model,
to $G_1=2.68$, $G_2=-1.84$ \cite{Jon93,Jon73}, recently quoted
values being $G_1=2.208$, $G_2=-0.556$ \cite{Gar93}.

The choice of the $\Delta$--propagator requires some caution.
In \cite{Tia78}, the Rarita--Schwinger form
\begin{equation} P_{\Delta}^{\mu \nu}(p)={\not p + m_{\Delta} \over
p^2-m_{\Delta}^2+i\Gamma m_{\Delta} } \left( g^{\mu \nu}-{1 \over 3}
\gamma^{\mu} \gamma^{\nu} - {2 \over 3} {p^{\mu} p^{\nu} \over m_{\Delta}^2}
-{\gamma^{\mu} p^{\nu} - p^{\mu} \gamma^{\nu} \over 3 m_{\Delta}^2}
\right), \label{delprop} \end{equation}
is used where the replacement $m_{\Delta} \to m_{\Delta}-i\Gamma /2$
is made in order to account for inelasticities due to pion production.
As we wish to use $P_{\Delta}$
in the far off--resonance region ($p^2 << m_{\Delta}$), we have to
take the energy dependence of the width $\Gamma \to \Gamma(q)$ into
account. For our purpose, we rely on the
parametrization of \cite{Faa83} requiring $\Gamma$
to vanish below the pion production threshold:
\begin{eqnarray} \Gamma (q_{\pi N}) &=& 0, \quad q^2 \leq 0,
 \quad q_{\pi N} \equiv \mid \vec q_{\pi N} \mid \nonumber \\
\Gamma (q_{\pi N})&=& 2 \gamma (q_{\pi N} R/m_{\pi})^3/(1+(q_{\pi N}
R/m_{\pi})^2), \quad
q_{\pi N}^2 > 0. \label{Brand} \end{eqnarray}
Here, $q_{\pi N}$ denotes the maximum momentum in the $\pi N$--subsystem
of the process $NN \to NN\pi$:
\begin{eqnarray} q_{\pi N}^2&=&
(s_{\pi N}-(m_{\pi}-m_N)^2)(s_{\pi N}- (m_{\pi}+m_N)^2)
/ 4 s_{\pi N}, \nonumber \\
s_{\pi N}&=& (\sqrt{s}-m_N)^2, \end{eqnarray}
$\sqrt{s}$ is the invariant energy in the $NN$--system
and $R$, $\gamma$ are adjustable parameters. Values of $\gamma = 0.71 MeV$,
$R=0.81$ lead to a resonance width $\Gamma=120$ MeV at $s_{\pi N}=1236$
MeV \cite{Faa83}. This value might be slightly modified by a variation of
$\gamma$. It has been shown \cite{Faa86}
that this parametrization is in good agreement with the experimental
$\delta_{33}$ phase as well as with a $\Delta$ self energy calculation
\cite{Ter87}.

$NN-N\Delta$ potential models differ in the choice of the
$\Delta$--coupling constants as well as in the definition of the meson
propagators and vertex form factors. We use these models as
a starting point for our calculation defining the $\Delta$--excitation
amplitude up to a normalization. In a reliable calculation,
the final results should not depend too much on the underlying
parameter sets provided they agree comparably well with the empirical
NN--data. We therefore compare two different recent
coupled channel models: The OBE--model of \cite{Ter87},
denoted as model A, and a coupled channel version of the Bonn
potential (model I of \cite{Mac89}, p. 353), denoted as model B.
For model A, we use the meson propagators
\begin{equation} P^{(\pi)}= {i \over  q^2-m^2_{\pi}}; \quad
P^{(\rho)}_{\mu \nu}= i {-g_{\mu \nu} + q_{\mu} q_{\nu}/m_{\rho}^2
 \over  q^2-m^2_{\rho}}. \label{mpropa} \end{equation}
The second term in $P^{(\rho)}$ drops out in the calculations of this and
the following section. The propagators of model B take the mass difference
between the nucleon and the $\Delta$ into account:
\begin{eqnarray} P^{(\pi)}= - i \left\{ {1 \over
2 \omega_{\pi}^2 } + {1 \over  2 \omega_{\pi} (m_{\Delta}-m_N+
\omega_{\pi}) } \right\}; \quad \omega_{\pi} \equiv \sqrt{m_{\pi}^2- q^2}
\, , \nonumber \\
P^{(\rho)}_{\mu \nu} = - i \left(-g_{\mu \nu} +
q_{\mu} q_{\nu}/m_{\rho}^2 \right) \left\{ {1 \over
2 \omega_{\pi}^2 } + {1 \over  2 \omega_{\pi} (m_{\Delta}-m_N+
\omega_{\pi}) } \right\}\, ,\label{mpropb} \end{eqnarray}
In writing down the Lorentz invariant propagators (\ref{mpropa}) and
(\ref{mpropb}) we have dropped the static approximation $q^2 \to -
\vec q\,^2$ of the potential models A and B. As we are far from the
pole in all our geometries and rescale our final amplitudes this has
very little effect on the end results.

For the calculations using model A, each
$NN\pi / NN\rho$--vertex is multiplied by a monopole cutoff
$\Lambda_{\pi / \rho}^2/(\Lambda_{\pi / \rho}^2 - q^2)$
and a dipole cutoff
$\left[ \Lambda_{{\scriptscriptstyle \Delta} \pi/\rho}^2/
(\Lambda_{{\scriptscriptstyle \Delta} \pi/\rho}^2
 + \vec q^2)\right]^2$
applies at the $N \Delta \pi / \rho$--vertices. Here again, we use
Lorentz invariant instead of static expressions. For model B, both the
$NN$-- and $N  \Delta$--vertices are
regularized by cutoffs $\left[ (\Lambda^2-m_{\pi/\rho}^2)/
(\Lambda^2-q^2) \right]^n$
where $n=1, \, 2$ for $\pi, \, \rho$ exchange respectively.

In general, the relativistic off--shell $\Delta$--particle is allowed
to propagate with both spin 3/2 and spin 1/2. This extra freedom is
reflected by additional off--shell terms in the interaction
Lagrangians; the more general chiral $\pi N \Delta$--vertex functions
corresponding to the propagator (\ref{delprop}) read \cite{Wit88,Ols75}:
\begin{eqnarray}
{\Lambda}_{\pi \scriptscriptstyle N \Delta}^{\mu} &=&
 {g{\pi \scriptscriptstyle N \Delta} \over m_{\pi} }
\Theta^{\mu \nu}(Z_{\pi}) q_{\nu} \, ; \quad
{\Lambda}_{\rho \scriptscriptstyle N \Delta}^{\mu \nu} =
 i {g_{\rho \scriptscriptstyle N \Delta} \over m_{\rho}}
\left( \not q \Theta^{\mu \nu} - \gamma^{\mu} q_{\alpha} \Theta^{\alpha \nu}
\right) \gamma_5 \, ,
\nonumber \\
\Theta^{\mu \nu}(Z_{\pi / \rho})&=& g^{\mu \nu} -(Z_{\pi / \rho}
+ {1 \over 2})
\gamma^{\mu} \gamma^{\nu}.
\label{lagoff}  \end{eqnarray}
A generalization similar to $\Lambda_{\rho \scriptscriptstyle N \Delta}$
(with parameter $Z_{\gamma}$) applies for the (leading) $G_1$--term
of equation (\ref{gamdel}). The $Z$--parameters
are not well determined either theoretically or experimentally. Whereas
the simple coupling scheme corresponding to $Z=-1/2$ is generally
preferred in potential models, Olssen and Osypowski \cite{Ols75} find
experimental values of $Z_{\pi}=0 \pm 1/4, \; Z_{\gamma}= 1/4 \pm 1/4$.
Note that the off--shell parameters do not affect a coupled channel
calculation at the OBE level but directly enter the Born amplitude for the
bremsstrahlung process where the $\Delta$ is off--shell.
We therefore rely on the parameters of Table 1 as fixed by NN--data
(to all orders of a coupled channel calculation)
but check the sensitivity of our results to a variation of $Z$.

This completes the necessary ingredients for the $\Delta$--excitation
part of the bremsstrah\-lung amplitude (\ref{deltamp}). With the
definitions of this chapter we have tried to incorporate
as much as possible of the experimental information
on the $N \Delta$ excitation channels in the NN--interaction which
have become available since the earlier Born calculations
\cite{Tia78}, \cite{Szy77}. On the other hand, the relativistic
approach allows us -- in contrast to a pure potential model calculation
\cite{Jon93} -- to extend the analysis to energies far beyond the
pion production threshold.

\begin{table}\centering
\begin{tabular}
      {@{~~~}c@{~~~}@{~~~~~~~~~~~}c@{~~~~~~~}c@{~~~~~~~~}
                             @{~~~~~~}c@{~~~~~~~}c@{~~~}}
  \\
        &
\multicolumn{2}{c}{\hspace{-1.5cm}{\large Model A \cite{Ter87}} }
&
\multicolumn{2}{c}{\large Model B \cite{Mac89}   }
                                       \\
  \\

{\mbox{\rule[-.40cm]{0cm}{0.9cm}}} &
\mbox{\rule[-.4cm]{0cm}{0.9cm}${g^2 \over 4 \pi}$}  &
\mbox{\rule[-.4cm]{0cm}{0.9cm}$\Lambda [{\rm MeV}]$}  &
\mbox{\rule[-.4cm]{0cm}{0.9cm}${g^2 \over 4 \pi}$}  &
\mbox{\rule[-.4cm]{0cm}{0.9cm}$\Lambda [{\rm MeV}]$}
  \\
 \hline     & & & &  \\
$NN\pi$
 & 14.16 & 1140  & 14.4 & 1800
\\
$NN\rho$
 & 0.43 & 1140  & 0.7 & 2200
\\

 & $\kappa =5.1$  &
 & $\kappa=6.1$   &
\\
$N\Delta \pi$
 & 0.35 & 910  & 0.35  &  920
\\
$N \Delta \rho$
 & 4.0  & 910  & 19.0  &  1140
\\
\end{tabular}
\medskip
\caption[]{Meson parameter of model A and B. For the definition of
form factors, propagators etc. see text. \label{Tab1} }

\end{table}

\section{Inclusion of the radiative $\omega / \rho$--decay}

We write the Lagrangian for the coupling of the $\omega$--meson to
the nucleons in analogy to equation (\ref{lag2}):
\begin{equation}
{\cal L}_{\omega NN} = -g_{\omega}\left(  \bar \psi  \gamma^{\mu}
\psi \omega_{\mu} + {\kappa_{\omega} \over 4 m_N }
\bar \psi \sigma^{\mu \nu} \psi (\partial_{\mu} \omega_{\nu}
- \partial_{\nu} \omega_{\mu}) \right), \label{lag5} \end{equation}
and parametrize the radiative decay vertices as in \cite{Gar93}
\begin{eqnarray}
{\cal L}_{\omega \pi \gamma} &=& g_{\omega \pi \gamma} \varepsilon_{
\nu \rho \sigma \delta} (\partial^{\sigma} A_{\gamma}^{\rho})
(\partial^{\nu} \pi^0) \omega^{\delta} \to \Lambda_{\rho \delta}^{
\omega \pi \gamma}(k,q) = -i g_{\omega \pi \gamma}
\varepsilon_{\nu \rho \sigma
\delta} k^{\sigma} q^{\nu} , \\
{\cal L}_{\rho^{i} \pi \gamma} &=& g_{\rho^{i} \pi \gamma}
 \varepsilon_{\nu \rho \sigma \delta} (\partial^{\sigma}
 A_{\gamma}^{\rho})
(\partial^{\nu} \pi^{i}) \rho^{i,\, \delta}; \quad i=0,+,-. \end{eqnarray}
With the meson propagators as in (\ref{mpropa}) and a vertex function
 $\Lambda_{\omega \scriptscriptstyle NN}^{\mu}$ as in eq. (\ref{vertices})
this yields for the diagram (c) of Fig. 1
\begin{eqnarray}
T^{(c)}_{\omega \pi \gamma}(p_1,p_2;p_3,p_4)
& = &  \bar u(p_3) \Lambda^{\omega NN}_{\mu}(p_3-p_1) u(p_1)
P_{(\omega )}^{\mu \delta}(p_3-p_1)
 \Lambda^{\omega \pi \gamma}_{\delta \rho}(k,p_4-p_2)
\epsilon_{\gamma}^{\rho} \nonumber \\ \mbox{} && \times \quad
  P_{(\pi)}(p_4-p_2)
\bar u(p_4) \gamma_5 u(p_2) \, , \label{radamp} \end{eqnarray}
and consequently (see equation (\ref{deltamp}))
\begin{eqnarray} T_{\omega \pi \gamma}^{pp\gamma} &=&
 T(p_1,p_2;p_3,p_4) + T(p_2,p_1;p_4,p_3)
\nonumber \\  && \mbox{} -
 ( T(p_1,p_2;p_4,p_3) + T(p_2,p_1;p_3,p_4)  ), \nonumber \\
T_{\omega \pi \gamma}^{np\gamma} &=&
 T(p_1,p_2;p_3,p_4) - T(p_2,p_1;p_4,p_3).
\label{omegamp}  \end{eqnarray}

The $\rho$--decay amplitudes are
obtained from $T_{\omega \pi \gamma}$ by interchange of the
masses and coupling constants and multiplication with the isospin
operator $\vec \tau_1 \cdot \vec \tau_2$. The matrix elements of this
operator yields factors of 1 and 0 for the neutral and charged
pion decay amplitude in $pp\gamma$. For $np\gamma$ the corresponding
values are -1 and 2.

The coupling constants $g_{\omega \pi \gamma}$ and $g_{\rho \pi
\gamma}$ are determined from the experimental decay widths of the
vector mesons \cite{Dat92}:
\begin{equation} \Gamma (\omega/\rho \to \pi \gamma)= {g_{\omega / \rho
\pi \gamma}^2 \over 96 \pi} {(m_{\omega/ \rho}^2-m_{\pi}^2)^3 \over
m_{\omega / \rho}^3} = \left\{ \begin{array}{r c l} 716 \pm 75 &
{\rm keV }&
{\rm for}\; \omega ,  \\ 121 \pm 31 & {\rm keV} &  {\rm for}
\;  \rho^0 , \\ 68 \pm 7 & {\rm keV} & {\rm for}
\;  \rho^{\pm} . \end{array} \right.  \nonumber \end{equation}

This leads to the numerical values
 \begin{equation}
{g_{\omega \pi \gamma}^2 \over 4 \pi}
=0.715 \cdot 10^{-3} m_{\pi}^{-2}; \quad
{g_{\rho^0 \pi \gamma}^2 \over 4 \pi}
=0.125 \cdot 10^{-3} m_{\pi}^{-2}; \quad
{g_{\rho^{\pm} \pi \gamma}^2 \over 4 \pi}
=0.070 \cdot 10^{-3} m_{\pi}^{-2}.
\label{cougam} \end{equation}

For the remaining coupling constants and cutoffs, we again follow the
rationale of the previous section and use a complete set that has been
successfully tested in pion--photoproduction \cite{Gar93}.
Thus we put
\begin{equation}
{g_{\rho}^2 \over 4 \pi} = 0.563; \quad \kappa_{\rho}=3.71; \quad
{g_{\omega}^2 \over 4 \pi} = 5.07; \quad \kappa_{\omega}=-0.12 \,.
\label{counuc} \end{equation}

These values
are in agreement with the quark model and the vector dominance assumption.
All form factors in this section are set to 1; the pion couplings
are taken from table 1. Note that a measurement of the vector meson
radiative decay contributions in the pion photoproduction process
determines the product of the couplings in eqs. (\ref{cougam})
and (\ref{counuc}). With the well--known empirical value of
$g_{\pi}$, the amplitude (\ref{radamp}) is therefore essentially fixed.

\section{Results}

As mentioned earlier, the amplitudes (\ref{deltamp}) and (\ref{radamp})
must be divided by an energy dependent scaling factor $g(T_{lab})$ in
order to compensate for neglecting higher order terms in the Born series.
{}From a comparison of our Born results using model A with the total
$\Delta$--absorption cross sections of \cite{Ter87} calculated in the
iterated coupled channel model, one finds a factor $g=1.65$ at lab energies
below 300 MeV. A calculation of the total $\Delta$--production cross
section and comparison with the experimental data at 800 MeV \cite{Hud78}
and 970 MeV \cite{Dmi86} suggests a weak energy dependence of $g$ so that
$g(730$ MeV$)=1.4$.
The latter value is in close agreement with the result of \cite{Sch94}.
As we have to compensate also for the the change from static to Lorentz
invariant propagators in eqs. (\ref{mpropa}), (\ref{mpropb}), we use
$g(280$ MeV$)=1.95$ and $g(730$ MeV$)=1.65$ respectively.

If we suppose the higher order corrections to be of equal importance
for both model A and B and for the $\omega/\rho$--decay amplitudes for
which no experimental comparison could be made, the model is completely
determined with the empirical values of $g(T_{lab})$ found above.
The rescaled amplitudes
(\ref{deltamp}) and (\ref{radamp}) can thus be coherently added
to the corresponding transition
matrices of the potential model \cite{Jet94} and the
soft--photon approximation \cite{Fea72,Fea79} so that the effects upon
both cross sections and spin observables can be studied. The relevant
formula for the observables considered can be found in the literature
(\cite{Lio72,Tia78}). For the potential model calculations we
shall focus on the domain around pion production threshold
where both correction effects should be maximum; the soft photon model
will be used to reanalyze the experimental data of \cite{Nef77}.

In principle, with all conventions carefully chosen, the formulas in
the last two  sections give the correct relative signs for the corrections
with respect to the leading (potential model or SPA) amplitudes. For an
independent check we combined our amplitudes (\ref{deltamp})
and (\ref{radamp}) with the pure one--pion exchange bremsstrahlung
amplitude (Fig. 1, diag. (a) with the internal $\Delta$--line replaced
by a nucleon line). If the exchanged pion is numerically put
on--shell ($q^0 \to \sqrt{\vec q^{\,2}+m_{\pi}^2}$), this is
equal up to a common factor to the pion--photoproduction amplitude
with and without intermediate $\Delta$ excitation. In a near--threshold
geometry ($\vec q^{\, 2} \sim 0$), the $\Delta$--amplitude gives
enhancement of the cross section by a few percent, in agreement with
Peccei's chirally invariant Lagrangian for $\pi^0 p \to p \gamma$
\cite{Pec69} whereas the $\omega/\rho$--decay amplitudes interfere
destructively, as in \cite{Gar93}.

The sign of the $\omega$--coupling constant given in eq. (\ref{cougam})
is consistent with the pion--photoproduction data \cite{Gar93}.
The sign of the much smaller radiative $\rho$--decay amplitude, however,
is still the object of controversies \cite{Sar89}.
We adopt here the sign convention of Gari et al.\ \cite{Gar76} where
the $\rho$--decay enhances the contribution of the $\omega$--decay
in $pp\gamma$; as the $\rho$--contribution is only about 2 $\%$
of the $\omega$ (compare the respective coupling constants), a switch in the
sign of $g_{\rho \pi \gamma}$ would not alter our conclusions.

\subsection{Potential Model Results}

All the calculations in this section are based on the inversion potential
to the Nijmegen-II NN--phase shifts used in Ref. \cite{Jet94}.
Using another realistic NN--potential would not change any of our results.
The basic features of the potential model can be essentially represented
in terms of two variables: the total energy of the process $T_{lab}$
as a crucial parameter for the relative size of the various
contributions in the amplitude, and the opening angles of the two
protons determining the maximum photon energy and thus the off--shell
signature of the process \cite{Jet94}. The two examples of Fig. 2
show the coplanar, exclusive $pp\gamma$ cross section $d^3 \sigma /
d \Omega_1 d \Omega_2 d \theta_{\gamma}$ and the analyzing power $A_y$
for the smallest and largest outgoing proton angle pairs measured in
the $T_{lab}=280$ MeV TRIUMF experiment \cite{Mic90}.
The plots show
that both $\Delta$--excitations and internal radiative decay contributions
are small for the cross section at small proton angles as well as for the
analyzing power $A_y$ at large proton angles. The $\Delta$--contribution
is the leading effect but tends to be partly cancelled by the radiative
decay contributions. The maximum net effect of both contributions can
increase the differential cross section by $15-20\%$ at medium photon
angles but amounts to only a few percent in the total cross section.

Among the examples of Fig. 2, only the 12.4$^0$, 14$^0$ analyzing power
shows a sizeable dependence on the parametrization of the $\pi$ and
$\rho$--exchange.
The choice of different electromagnetic couplings, e.\ g.\ $G_1=2.68$,
$G_2=-1.84$, increases the $\Delta$--contribution in the cross sections
by about 20 $\%$ which translates to an enhancement of the complete
$pp\gamma$ cross section by about 5 $\%$ at maximum (Figure 3).
Conversely, the
choice of $Z_{\pi}=Z_{\gamma}=-1/4$ \cite{Gar93} reduces the
size of the $\Delta$--contributions by roughly $20\%$. These two choices for
the couplings and off--shell parameters are likely to be rather extreme.
Moreover, because of its small relative size, a rescaling error in the
$\omega/\rho$--amplitude would not affect much the end results. One
might therefore conclude that the corrections are
reasonably well determined in our model.

In order to obtain a more general picture we calculate the double
differential cross section $d \sigma/ d \Omega_1 d\Omega_2 =
\int d  \theta_{\gamma} ( d \sigma / d \Omega_1 d \Omega_2
d \theta_{\gamma})$ and represent its relative
enhancement due to the correction
terms as a function of the symmetric angle $\theta=\theta_1=\theta_2$
(Figure 4). Note that small $\theta$ values correspond to a suppression of the
meson four momentum transfer ($q \to 0$) whereas for large $\theta$,
the elastic limit $k\to 0$ is reached. In both cases, the amplitudes
(\ref{deltamp}) and (\ref{radamp}) are suppressed: they are maximum in
the medium proton angle region $\theta \sim 20^0$.
The relative size of the corrections dies down as the energy
decreases and reaches only  about 3$\%$ at $T_{lab}=200$ MeV.

In Figure 5 the analysis of Fig. 2 is repeated for the
$np\gamma$--observables. Here, the isospin factors yield strong
destructive interference between the various Feynman graphs so that
both the $\Delta$ excitation and the radiative decays become
negligible corrections to the potential model cross section. The same
result holds for the $np\gamma$--analyzing power.
We thus confirm the result of \cite{Boh77} that $\Delta$--corrections
to the $np\gamma$--amplitude are weak.

\subsection{730 MeV (SPA) Results}

Two typical geometries of the 730 MeV $pp\gamma$--experiment reported in
Ref. \cite{Fea79} are shown in Fig 5. One proton is emitted at $\theta_1
=50.5^0, \, \phi_1=0^0$ and the polar angles of the photon are
$\theta_{\gamma}=67^0, \, \phi_{\gamma}=179^0$ for counter G7 and
$\theta_{\gamma}=54^0, \, \phi_{\gamma}=131^0$ for counter G10.
The solid curves have been obtained
with the soft photon approximation of \cite{Fea72}. Contributions in
${\cal O}(k)$ are only partly included in SPA; the corrections considered
here are thus specific candidates for the missing ${\cal O}(k)$ effects.

As expected the corrections are negligible for small $k$ but increase with
the photon energy. Analogous to the 280 MeV examples, there is
cancellation between
the $\Delta$-- and $\omega/\rho$--effects. The actual size is again
geometry dependent but is limited to about 20$\%$ of the SPA cross section
at $k=150$ MeV. A variation of the electromagnetic coupling constants and
off--shell parameters within the experimental limits yields effects
analogous to those shown in Fig. 3 and leaves a freedom of a few percent
in our final $k=150$ MeV results. Correspondingly, the results are not
much altered by interchange of model A and B. We conclude that a
$\Delta$--amplitude consistent with experimental results can only resolve
a small part of the discrepancy between the SPA and the 730 MeV
$pp\gamma$--results at photon energies above $\sim$ 100 MeV.

\section{Concluding Remarks}

We have evaluated the $\Delta$--excitation and radiative $\omega/\rho$--decay
corrections to proton--proton and neutron--proton bremsstrahlung for the
energy range up to $T_{lab}\sim 1$ GeV. This was done by calculating
the relativistic Born amplitudes and adding them to potential model and SPA
amplitudes. The Born amplitudes were  normalized by calculating
$\Delta$--production and absorption cross sections in the same model and
fitting to experimental data and coupled channel predictions.

Of the two processes considered, the $\Delta$--excitation dominates but is
generally partly compensated by the radiative decay contributions.
Both corrections together increase the 280 MeV $pp\gamma$ integrated
cross section $d^2 \sigma / d \Omega_1 d \Omega_2$
by an amount depending essentially on the proton opening angles $\theta_i$
and reaching a maximum of roughly 7.5$\%$ at $\theta_1 \sim
\theta_2 \sim 20^0$.
The relative effect on $d^3 \sigma / d \Omega_1 d \Omega_2 d \theta_{\gamma}$
is maximum for photon emission around $\theta_{\gamma}=90^0$ and is
typically $\leq 20\%$.

In the 730 MeV Rochester geometry, the corrections become relevant for
photon energies above $\sim$ 100 MeV but are not big enough to
complement the soft photon approximation and fit the experimental data
\cite{Fea79}.

We have taken care to study and discuss the limits and uncertainties of
our model. The results have been obtained on the basis of two different
parametrizations of the $\Delta$--excitation (model A and B of Table 1)
but show little sensitivity to the underlying coupled--channel model.
The fact that the $\Delta$--excitation cross sections of
Model A and B are nearly identical, despite the rather massive differences
in the values of the $\rho$--couplings and the choice of form factors and
propagators, is reassuring for our approach but
corresponds to the result obtained earlier that rather different potential
models, once they fit the NN--scattering data,
yield very similar bremsstrahlung results \cite{Jet94} and is
disappointing if one hopes to understand the underlying physical reaction
mechanism from $NN\gamma$ measurements.

The experimental uncertainty in the radiative $\Delta$--decay constants
and the off--shell freedom of the $\Delta$ suggest a theoretical error of
about $\pm 20\%$ in the final correction amplitudes, i.\ e.\
up to $\sim \pm 4\%$ in the total (280 MeV) $pp\gamma$ cross section.
We should mention that the discrepancies between the $pp\gamma$
predictions of different NN--potential models, which most bremsstrahlung
experiments in the past were designed to isolate, are of the same order
of magnitude. Given the present experimental status
of nucleon--nucleon bremsstrahlung and the theoretical ambiguities mentioned,
it is not possible to use $NN\gamma$--data for putting limits on
the parameters entering our correction amplitudes.

In view of the similarities of our calculation with previous ansatzes,
a comprehensive comparison of the results seems worthwhile. First note
that we have eliminated the uncertainty with respect to the relative sign
of the relativistic corrections stated by Kamal and Szyjewicz
\cite{Kam77} through a comparison with the
pion--photoproduction process. Our $\omega/\rho$--decay amplitude
extends the amplitude of \cite{Kam77} by including the
$\omega$--tensor coupling terms and the $\rho$--decays and is, if
we divide by $g(T_{lab})$, slightly smaller than given there.

The discrepancy of our result is more serious for the 730 MeV
$\Delta$--excitation calculation where the previous authors lack a reliable
model for the NN--channel $pp\gamma$--amplitude. The authors of
\cite{Tia78} therefore
essentially fit the $\Delta$--excitation part to the experimental
$pp\gamma$ data. In \cite{Szy77}, the suppression of the amplitude
which we simulate by appropriate form factors is
neglected. In both cases, the pure $\Delta$--cross section becomes bigger
than ours by more than a factor of 10 and would be in clear
contradiction to the 800 MeV $\Delta$--production data of, e.\ g.\
\cite{Hud78}. We also stress the importance of  taking interference
terms correctly into account.

The dispersion theoretic approach used in \cite{Boh77} to estimate the
role of the $\Delta$--resonance in a OPE--$np\gamma$ calculation is
rather different from our model. Correspondingly, the results agree
only in relative size.

Finally, a comparison with the coupled--channel $\Delta$--results of
\cite{Jon93} in the low energy region shows good agreement for the
cross sections and even for the analyzing powers $A_y$. This is
encouraging for the feasibility of our method as well as for the
various extensions made.

\bigskip

We wish to thank Dr. S. Scherer for valuable discussions.
This work was supported in part by a grant from the Natural Sciences and
Engineering Research Council of Canada.

\pagebreak

\pagebreak
\normalsize
\begin{figure}
\caption[]{\label{feynman} $\Delta$--excitation and internal
radiation processes considered in this text.}
\end{figure}

\begin{figure}
\caption[]{ \label{fig2} Coplanar $pp\gamma$ exclusive cross sections
and analyzing powers at $T_{lab}=280$ MeV and for the smallest and
largest proton angle pairs $\theta_1, \, \theta_2$ of the TRIUMF
experiment \cite{Mic90} (the datapoints shown are rescaled with a factor
of 0.67 as in this reference). The curves denote the pure potential
model of Ref. \cite{Jet94} (solid), potential model plus pure
$\omega / \rho$--decay
(long dashed), and the full model according to model A (dotted) and
B (dashed--dotted). }
\end{figure}

\begin{figure}
\caption[]{ \label{fig3} Typical uncertainties of the $\Delta$--contribution.
The curves show the pure potential model of Ref.
\cite{Jet94} (solid) and the full calculation according to model A
with standard parameters $G_1=2.208$, $G_2=-0.278$, $Z_{\pi}=
Z_{\gamma}=-1/2$
(long dashed). The upper bound of the shaded area in the cross section and
lower bound in $A_y$ corresponds to larger electromagnetic
couplings $G_1=2.68$, $G_2=-1.84$. The lower bound in the cross section
and upper bound in $A_y$ is calculated with larger $Z$--parameters
$Z_{\pi}=Z_{\gamma}=-1/4$.}
\end{figure}

\begin{figure}
\caption[]{ \label{fig4} Relative change of the integrated, coplanar
$pp\gamma$ cross section
$d^2 \sigma / d \Omega_1 d \Omega_2$(with corrections)/$d^2
\sigma / d \Omega_1 d \Omega_2$(without corrections) at
280 MeV (dashed), 200 MeV (dashed--dotted) and 157 MeV (dotted)
for parameter set A as a function of the symmetric proton
angles $\theta_1=\theta_2$. }
\end{figure}

\begin{figure}
\caption[]{ \label{fig5} Coplanar $np\gamma$ exclusive cross section
and analyzing power at $T_{lab}=280$ MeV.
The curves denote the pure potential model of Ref.
\cite{Jet94} (solid), plus pure $\omega / \rho$--decay (long-dashed),
and the full model according to model A and B
(dashed--dotted and dotted).}
\end{figure}

\begin{figure}
\caption[]{ \label{fig6} 730 MeV $pp\gamma$ measurements and soft photon
approximation in two typical geometries (see \cite{Fea79} (1979) and text).
The curves denote the pure SPA results (solid), plus pure $\omega /
\rho$--decay (long-dashed), and the full model according to model A (dotted).}
\end{figure}

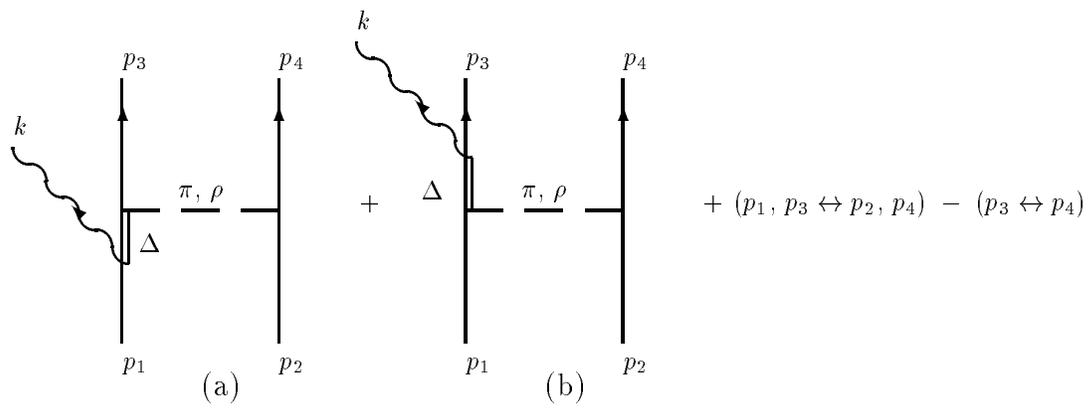
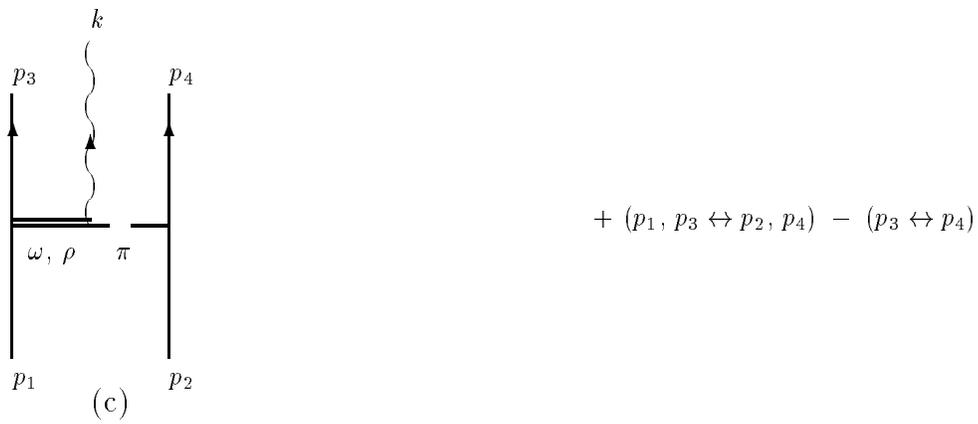
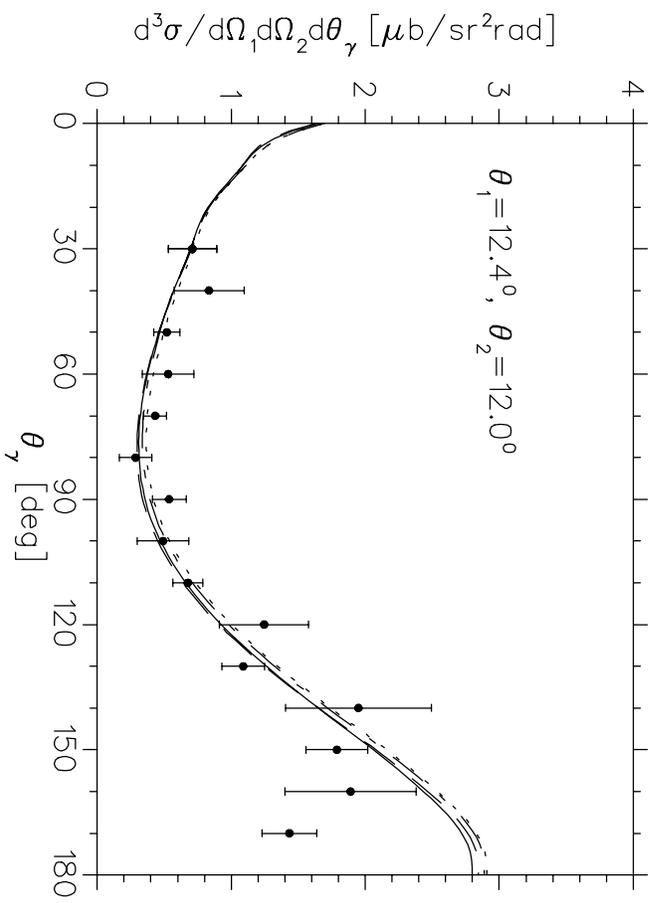
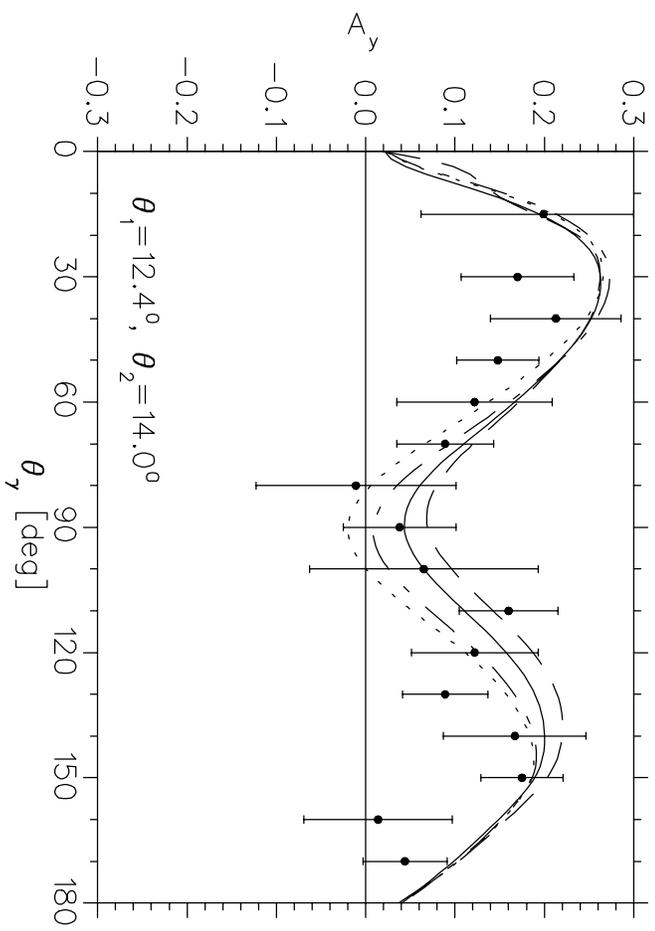
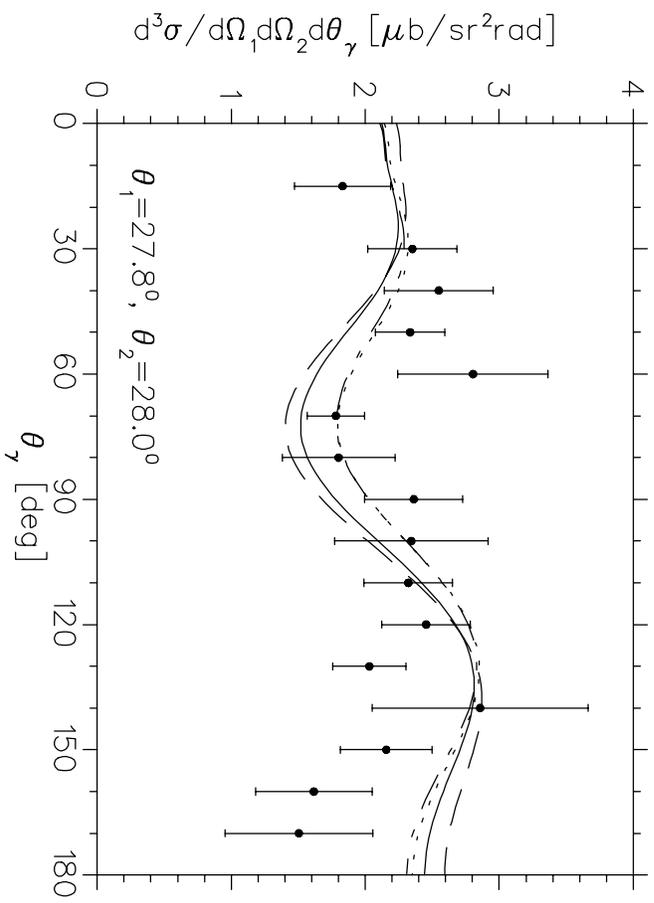
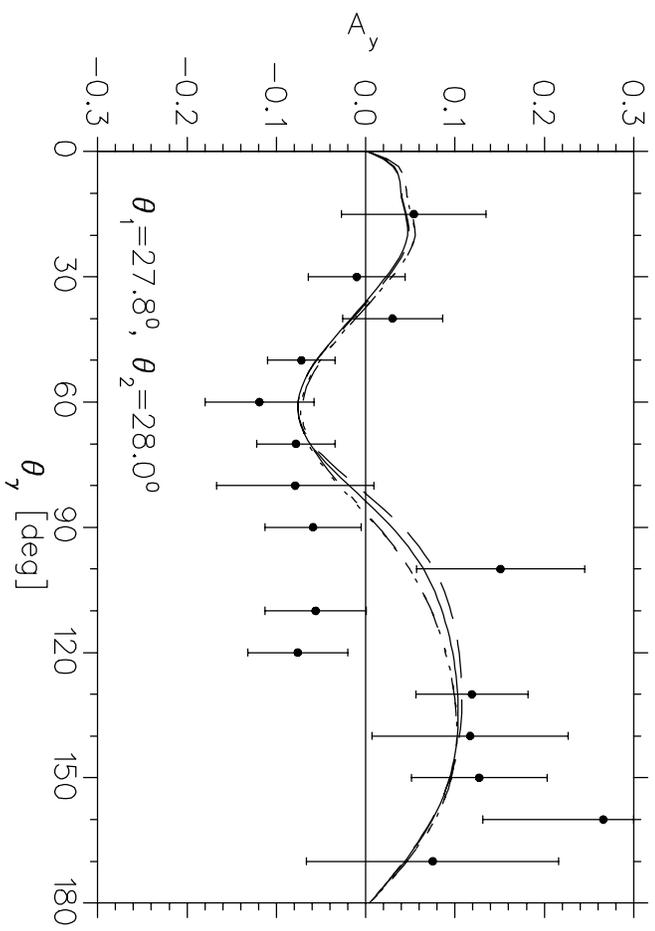
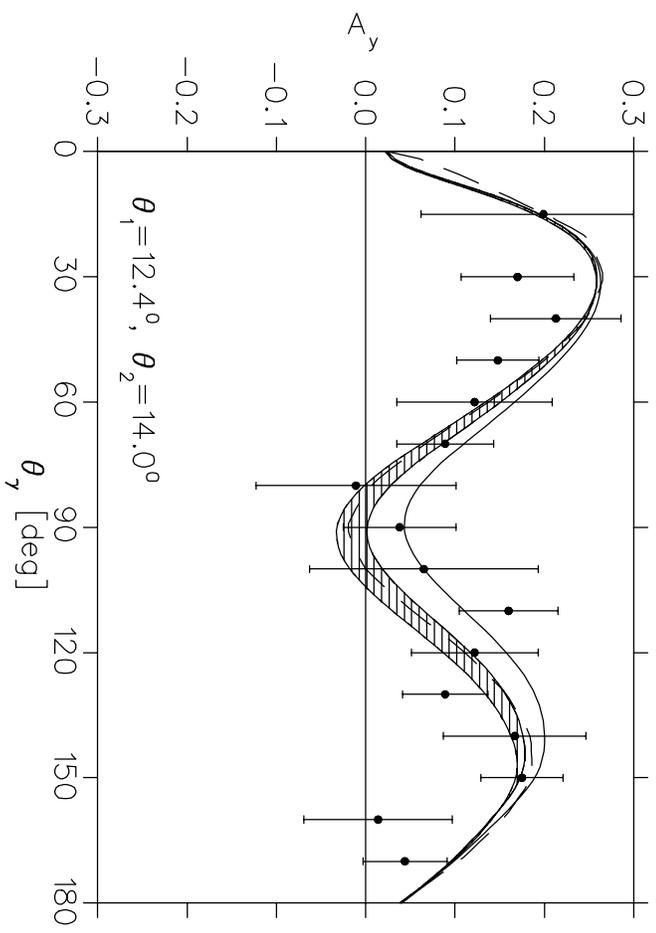
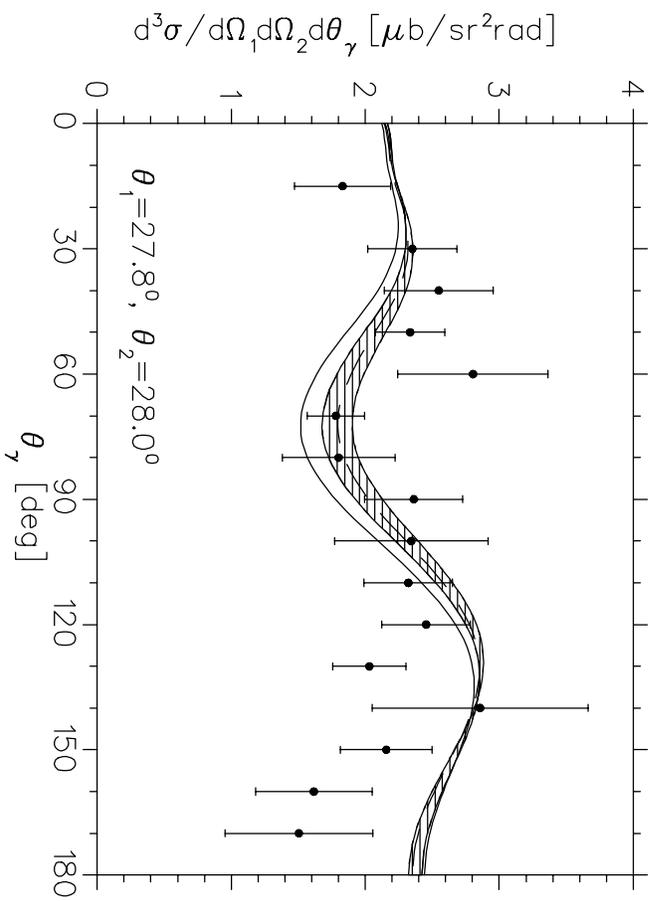
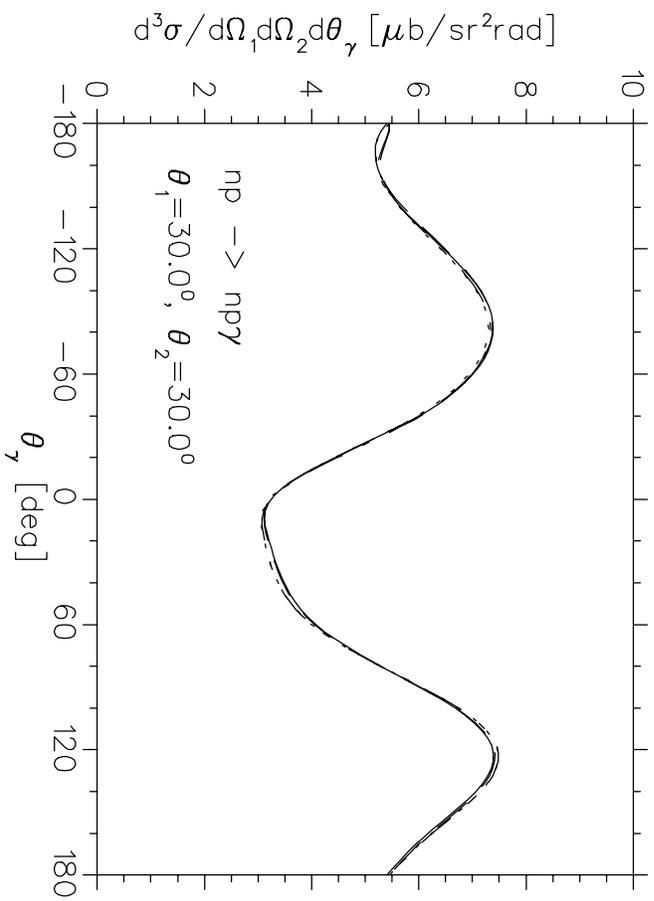
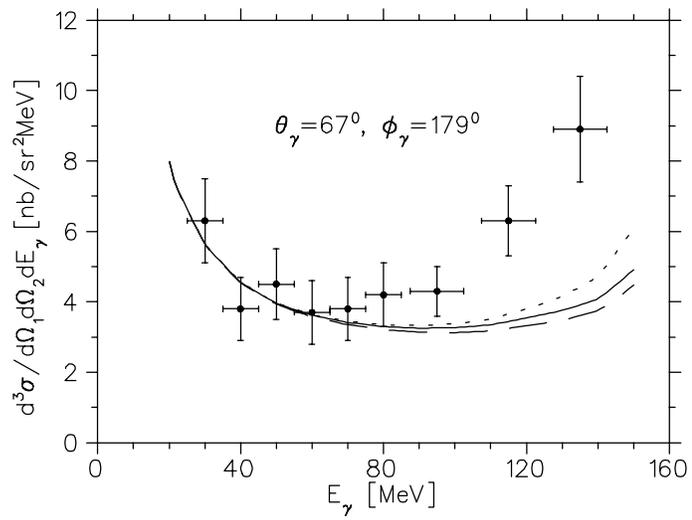
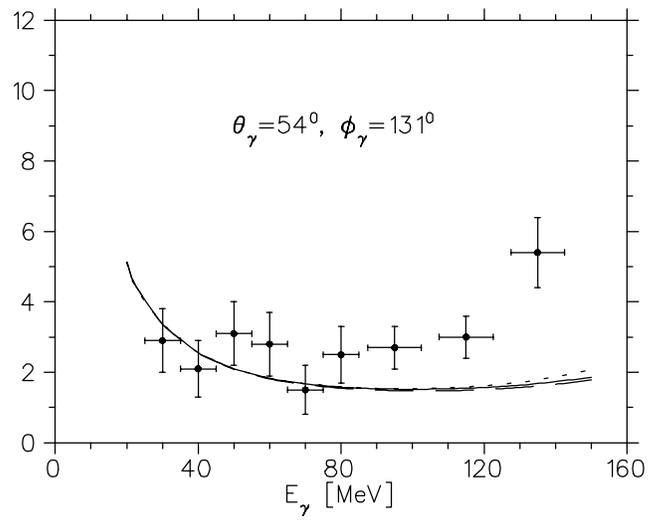
\end{document}